# Modelling of the NBI contribution to the neutron energy spectra for the ITER Vertical Neutron Camera

**T. Kormilitsyn\*, G. Nemtsev, R. Rodionov, Yu. Kashchuk, and D. Portnov**

*Project Center ITER*
*Kurchatov sq. 1 bld. 3, 123182 Moscow, Russia*
*E-mail*: t.kormilitsyn@iterrf.ru

ABSTRACT: The ITER Vertical Neutron Camera (VNC) is a multichannel collimator system designed to measure the neutron emissivity profile and spectra. The energy spectrum measurements of neutrons made by the VNC will contribute to the reconstruction of the ion temperature, fuel ratio, and fast ions' distribution function. In this paper, the results of neutron spectra calculations for the ITER VNC are presented. The developed software allows for the simulation of particle spectra resulting from interaction of suprathermal ions with thermal background plasma. The method relies on direct reaction rate calculation and explicit fusion reaction kinematics modelling. The assessment of the D-T neutron spectra is carried out for D-NBI heated baseline D-T ITER operation scenarios. It is shown that neutron flux resulting from fast ("beam") D with thermal T ion interaction dominates at $E_n > 16$ MeV part of the spectra. Thus, it is possible to assess fuel ratio based on calculated "thermal-thermal" and "beam-thermal" neutron energy spectra. An ability to observe sawtooth-induced mixing via VNC is studied. Lower energy boundary for the observable part of the suprathermal ion distribution is derived.

KEYWORDS: fast ion behavior, fast ion VDF, neutron spectroscopy, plasma diagnostic, plasma instability, plasma heating

**Contents**

## 1. Introduction

The study of fast ion behaviour in reactor conditions is among the major goals of the International Thermonuclear Experimental Reactor (ITER) project. Additional heating by neutral beam injection (NBI) creates population of suprathermal ions with anisotropic distribution in velocity space. The neutron energy spectrum measurements made by the Vertical Neutron Camera (VNC) can contribute to reconstruction of the following plasma parameters: fuel ratio, ion temperature, and fast ions' distribution function in combination with the radial measurements performed by radial neutron camera (RNC) and high-resolution neutron spectrometer (HRNS) diagnostics. It will help to monitor the consequences of instabilities which cause the redistribution of fast ions in the plasma and assess their impact on plasma heating and current drive. Assessing the capabilities of diagnostics at various phases of ITER operation is an essential part of ITER research planning. To investigate the required accuracy and the resolution of the measurements of the VNC in deuterium-tritium (DT) ITER scenarios, we have developed a computational module. The developed module enables simulation of the anisotropic spectra of neutrons originated from interactions between suprathermal (beam) and thermal (background) particles. This module uses input profiles of plasma parameters, plasma and collimator geometry, fast ion distribution function. The module has the interface compatible with the ITER scenario database and, thus, can be used in the simulations of the ITER plasma evolution.

## 2. Modelling setup

Each channel of the VNC features a diagnostic unit with a diamond detector that is used for neutron spectra measurements. All VNC lines of sight (LOS) are located in poloidal plane. For comparison we have selected an additional LOS aimed at the plasma tangentially in the equatorial plane. Figure 1 serves as an illustration of these lines of sight, with ITER poloidal cross-section on the left and ITER equatorial cross-section on the right.

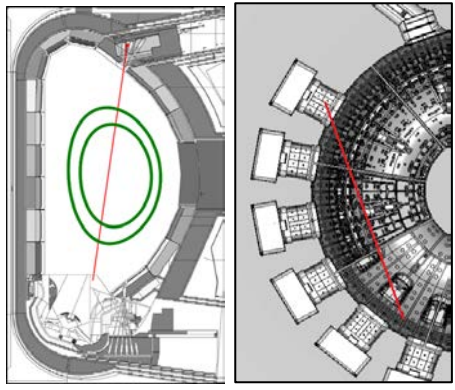

**Figure 1.** (—) Illustration of the lines of sight selected for this research. ITER Vertical Neutron Camera LOS on the left, tangential LOS – on the right.

To account for suprathermal ion population presence in the plasma, the fast ion distribution $f_{fast}(r, v, \cos\mu)$, with, $v = |\vec{v}|$, $\cos\mu = \vec{v} \cdot \vec{B}/(v|\vec{B}|)$, was simulated by ASTRA Fokker-Plank solver 2D in velocity space [1] for profiles expected in the ITER H-mode operation. The velocity distribution function displays significant anisotropy (Figure 2). The majority of fast ions have a pitch-angle close to the angle between the tangential LOS and the plasma magnetic surfaces.

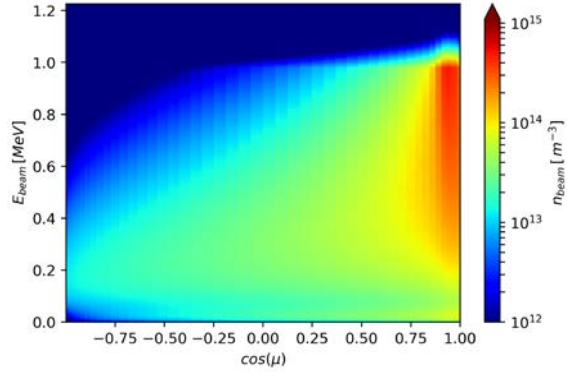

**Figure 2.** Fast ion velocity distribution function: $f_{fast}(r, v, \cos\mu)$ illustrated for $r = 0$ (plasma core).

Collimator length and typical diamond dimensions (~4x4x0.5 mm) allow us to use a simplified collimator geometry limited to collimator length and aperture size to calculate collimated 'beam-thermal' part of the observed spectra. For this a direct reaction rate calculation approach was selected. It employs 'beam-target' approximation to simplify the calculation. Assuming the background tritium ion is a rest:

$$S_{DT}(E) = \int_0^L n_T n_D v \sigma(v) f_{fast}(l, \mathbf{v}) \delta(E - E_n) \Delta\Omega dl d^3\mathbf{v},$$

here $n_{D,T}(l)$ - plasma ion densities, $v$ – beam ion velocity, $f_{fast}$ - is a fast ion distribution function, $l$ – distance along the LOS, $\sigma(v)$ - D-T reaction cross-section, $E_n(\mathbf{v}, \theta)$ – neutron energy derived from reaction kinematics [2], $\theta(l)$ - angle between LOS and magnetic surface.

Energy spectra part of neutrons originating from the background plasma ('thermal-thermal') was calculated via MCNP [3] with 25 keV of ion temperature incorporated in the source definition. The ITER C-Model release 180430 issued on 30/04/2018 model was used to model the 'thermal-thermal' VNC LOS spectrum. Figure 3 illustrates the UNVC MCNP model.

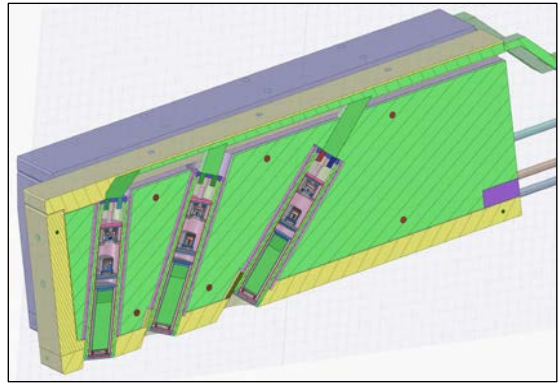

**Figure 3.** Upper Vertical Neutron Camera assembly cross section.

The spectrum for tangential LOS was estimated based on that of VNC. Both lines of sight are assumed to be directed at plasma core and both lines of sight employ miniature diamond detectros with sufficiently long collimators. The plasma density profiles are practically flat, thus, the 'thermal-thermal' part of the spectra depends mainly on the LOS length and solid angle:

$$S_{TT} = \int_0^L n_T n_D <\sigma v> \Delta\Omega dl$$, here $S_{TT}$ - observed 'thermal-thermal' spectrum, $n_{D,T}$ - plasma ion densities, $<\sigma v>$ - D-T reaction rate, $\Delta\Omega$ - collimator solid angle, $l$ – distance along the LOS.

The neutron flux observed by VNC LOSs was modelled for the ITER baseline DT scenario with $P_{fus}$ = 500 MW, 33 MW of the $D^0$-NBI heating at $E_{NBI}$ = 1 MeV, with the shapes of sources of $D_{th}$-$T_{th}$ neutrons and $D_{beam}$-$T_{th}$ neutrons (figure 4).

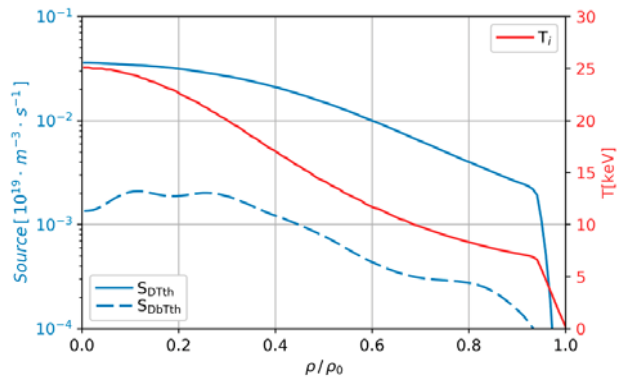

**Figure 4.** (—) 'thermal-thermal' (solid line) and 'beam-thermal' (dashed line) neutron source profiles and average ion temperature profile (—) versus normalized toroidal flux for ITER baseline D-T scenario. Profiles generated in ASTRA Fokker-Plank Solver.

## 3. Results and discussion

### 3.1 Sawtooth emulation

To assess the effect of fast ion velocity distribution function (VDF) anisotropy, we compared spectra calculated for selected Upper-VNC LOS and the tangentially aligned LOS.

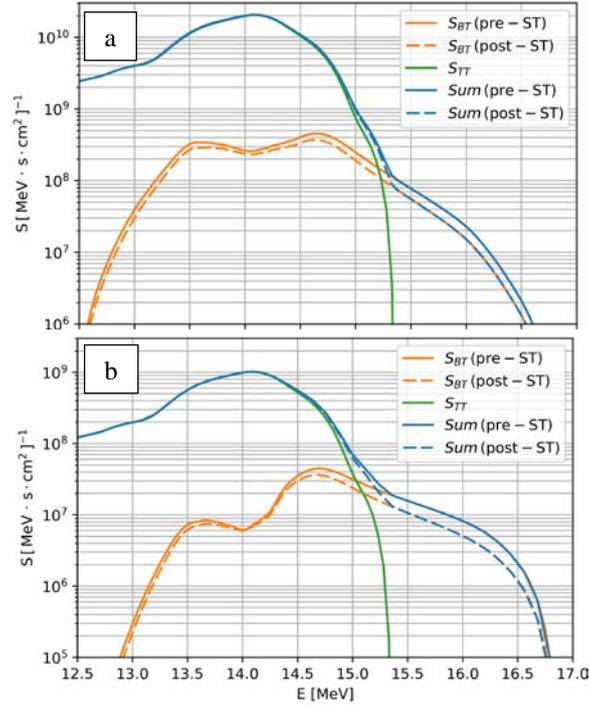

**Figure 5.** (—) $S_{TT}$ or 'thermal-thermal' part of neutron energy spectra, (—) $S_{BT}$ or 'beam-thermal' part before (solid line) and after (dashed line) the modelled redistribution, (—) total collimated spectra calculated for (a) – VNC LOS, (b) – tangential LOS.

Additionally, a model of fast ion density loss was employed to assess the ST-mixing detection capability. Redistribution was modelled as a fast ion density drop:

$$f_{fast}(r,v,\cos\mu) = f_{fast}(r_{max}/2, v, \cos\mu), \text{ in the central zone: } r \in [0; r_{max}/2].$$

Similarly to tangential LOS, the fast ion population influence on UVNC neutron energy spectra is visible at $E_n \in [15.5, 17.0]$ MeV (see fig. 5). Compared to the results obtained for tangential LOS [4] the redistribution is less visible. Define the 'beam-thermal' visible tail amplitude as the average spectrum in the [15.5, 16.0] MeV region and the 'thermal-thermal' peak amplitude as the max spectrum in the [13.5, 14.5] MeV region. Then 'beam-thermal' to 'thermal-thermal' amplitude ratio is lower for VNC (~0.2%) than that of tangentially aligned LOS (~0.8%) due to fast ion VDF anisotropy.

In both cases neutron energy spectra peak integration demonstrates ~3÷5% of 'beam-thermal' to 'thermal-thermal' neutron ratio. This result is in good agreement with the work by G. Ericsson [5].

### 3.2 NBI $E_{max}$ variation

We conducted the evaluation of fast ion population part responsible for the visible part of 'beam-thermal' neutron spectrum tail. The evaluation was done by beam maximum energy variation which was modelled as the energy axis remapping: given $E_{max} = m(v \cdot x)^2/2$, we assume that $f_{fast}(r, v \cdot x, \cos\mu) = f_{fast}(r, v, \cos\mu)$, where $x \in [0.1; 1.0]$.

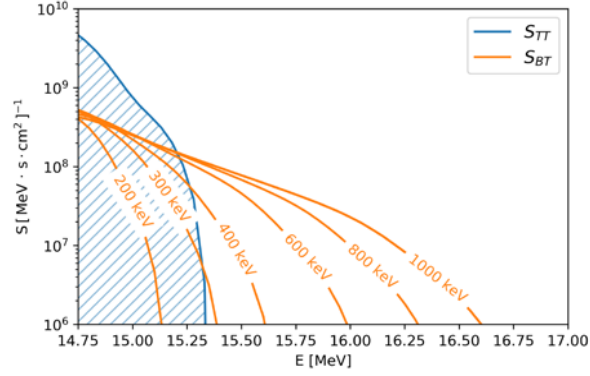

**Figure 6.** (—) $S_{tt}$ or 'thermal-thermal' part of neutron energy spectra, (—) $S_{bt}$ or 'beam-thermal' part calculated for various maximum beam energy.

Beam $E_{max}$ scaling shows that the 'beam-thermal' 'tail' visible by the UVNC (given the energy resolution of ~1% [6]) allows observing the 500+ keV part of the fast ion population (see fig. 6). The observation of the high-energy part of the population allows for heating efficiency assessment through beam slowing down time.

For baseline D-T scenario of ITER operation the typical timescale for NBI ion thermalization is 0.1-1 s [7]. This beam slowing down timescale allows for sufficient spectra sampling according to the ITER VNC measurement requirements to a time resolution of 10 ms.

The calculation shows the UVNC diagnostic system is able to assess the beam slowing down time directly after NBI shutdown. The opposite action can be used to evaluate the employed fast ion distribution model via externally supplied density and temperature parameters.

### 3.3 Fuel ratio measurements

Fitting the 'beam-thermal' neutron energy spectra component based on the experimental data ($S_{TT}$ & $S_{BT}$ on figure 7) and the analytical model allows for fuel ratio measurements via VNC.

The following equation can be applied:

$$n_t/n_d \approx \frac{(S_{BT})^2 <\sigma v>_{TT}}{S_{TT}(<\sigma v>_{BT})^2 (\int_0^L \frac{\partial n_{d_b}}{\partial l} dl)^2}$$

, where $n_{d_b}$ is fast ion density along the detector LOS, $S_{BT}$ is the observed 'beam-thermal' spectrum, $S_{TT}$ is the observed 'thermal-thermal' spectrum, $<\sigma v>$ - the corresponding reaction rates.

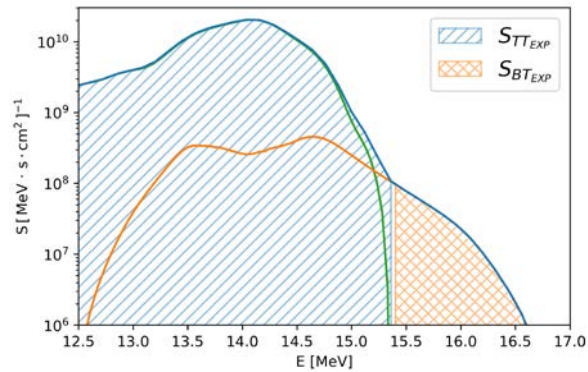

**Figure 7.** (—) $S_{TT}$ or 'thermal-thermal' part of neutron energy spectra, (—) $S_{BT}$ or 'beam-thermal' part. Filled areas illustrate the experimentally obtainable spectra parts.

The beam distribution function for the evaluation can be calculated via ASTRA or NUBEAM software. A similar method was demonstrated for the T-NBI shots data acquired during the JET DTE1 campaign [8]. The opposite action can be used to benchmark the employed fast ion distribution model via externally supplied fuel ratio value.

## 4. Conclusion and outlook

The NBI heating provides an explicit impact on the observed collimated neutron energy spectra which, in turn, provides opportunities to derive information on plasma fuel ratio, beam slowdown time, fast ion distribution function anisotropy, and redistribution resulting from MHD instabilities. The conducted calculation gives an estimation of the ITER VNC capabilities in terms of deriving of the discussed plasma parameters.

The next considerable step is benchmarking the calculation method versus other codes (ex. DRESS-code [9]). Upon calculation of the fast ion VDF for several time points, we plan to test the $\tau_{se}$ measurement accuracy and $T_e$ assessment procedure. Further implementation of the detector response function in the calculation will provide additional assessment of the required dynamic range and energy resolution.

## Acknowledgments

The author expresses gratitude to A. Polevoi of ITER Organization for providing the fast ion distribution model. Current work is supported with Russian Federation State Contract H.4a.241.19.19.1009 of December the 26[th] 2018 and conducted using the adaptation of the C-lite MCNP models which were developed as a collaborative effort between the FDS team of ASIPP China, University of Wisconsin-Madison, ENEA Frascati, CCFE UK, JAEA Naka, and ITER Organization.